\documentclass[aps,prl,nofootinbib,superscriptaddress,10pt,onecolumn]{revtex4-2}
\usepackage[utf8]{inputenc}
\usepackage{graphicx}
\usepackage{subcaption}
\usepackage{amssymb,amsmath}
\usepackage{xcolor}
\usepackage{float}
 \usepackage{caption}

\def\be{\begin{equation}}
\def\ee{\end{equation}}
\def\bea{\begin{eqnarray}}
\def\eea{\end{eqnarray}}

\def\s{{\rm s}}
\def\yr{{\rm yr}}
\def\cm{{\rm cm}}

\def\gev{{\rm GeV}}

\begin{document}

\title{A Strategy for Identifying Periodic Sources Contributing to the Galactic Center Excess}

\author{Eric J.~Baxter}
\affiliation{Institute for Astronomy, University of Hawai`i, 2680 Woodlawn Drive, Honolulu, HI 96822, USA}

\author{Jason Kumar}
\affiliation{Department of Physics and Astronomy, University of Hawai'i, Honolulu, HI 96822, USA}

\begin{abstract}
The origin of the Galactic Center gamma-ray excess (GCE) has not been conclusively determined after over a decade of careful study.  The two most widely discussed possibilities are a population of millisecond pulsars (MSPs), and annihilation of dark matter particles.  In contrast with annihilating dark matter, MSPs are expected to produce periodic emission.  We show that even though the number of photons contributing to the excess is small, there is potentially sufficient information in the data from {\textit Fermi} to detect a periodic MSP signal.  Such a detection would definitively prove that at least some fraction of the excess is due to MSPs.  We argue that this conclusion is robust to potential timing perturbations of the gamma-ray photons, such as those due to Earth's orbit, even if the number of parameters that must be used to model the perturbations is $\sim 7$. 

\end{abstract}

\maketitle

{\it Introduction.}  One of the most intriguing current puzzles in astrophysics is the origin 
of the Galactic Center excess (GCE) of GeV-range photons seen in {\textit Fermi}-LAT 
data~\cite{Goodenough:2009gk,Hooper:2010mq,Abazajian:2012pn,Fermi-LAT:2015sau}.   There have been a 
wide variety of approaches to determining if the GCE arises 
from dark matter (DM) annihilation,  millisecond pulsars (MSPs)~\cite{Abazajian:2010zy} or perhaps some other source.  
One such approach has been 
to search for non-Poisson fluctuations in the photon counts from the GC~\cite{Lee:2015fea}, which 
may be expected if some pixels host relatively rare but bright MSPs, which 
produce several detected photons.  Simple estimates indicate that, in optimistic 
scenarios, $\sim 10$ photons in a pixel in the $1.9 - 12~\gev$ energy range might arise from a single pulsar whose 
luminosity is just below the {\textit Fermi}-LAT point source threshold, with $\sim 10 - 50$ photons
arising from other sources~\cite{Lee:2015fea}  
in a pixel with size of roughly $0.5^{\circ}$.  

We consider whether photon timing information can provide additional discriminating power between MSPs and dark matter annihilation. In particular, since MSPs produce photons directed at Earth on a characteristic  $\sim 10^{-3}~\s$ timescale, while {\textit Fermi} has a timing resolution of 
order $10^{-6} \s$~\cite{Fermi-LAT:2021wbg}, one 
might hope that {\textit Fermi} could detect a telltale periodicity in photon arrival times in 
pixels from which a large excess of photons are seen.  
Such periodicity would not be be expected from dark matter annihilation, or from mismodeled astrophysical backgrounds.  In this Letter, we consider in detail the prospects for this strategy.

A key advantage of this search strategy for exploring the nature of the GCE is that it does not 
rely on having correctly modeled the amplitude, energy dependence, or spatial distribution of the various 
astrophysical or exotic gamma-ray emission processes.  Instead, one essentially analyzes the sum of 
all gamma-ray emission processes within a pixel, and determines if there is a component which is 
periodic in time.  As such, this strategy is immune to the difficulties in background modeling which 
plague some other approaches to studying the GCE.  The downside is that, although this strategy can 
potentially determine if a significant fraction of the GCE arises from bright MSPs, it cannot 
distinguish a contribution from numerous faint MSPs producing photons which are not correlated in 
time with each other.  
The time of arrival at {\textit Fermi}-LAT of gamma rays from identified MSPs has previously been used 
to constrain gravitational waves~\cite{Fermi-LAT:2022wah} and wave dark matter~\cite{Xia:2023hov,Luu:2023rgg}, 
using the MSPs as a pulsar-timing array.  In this analysis, 
by contrast, we consider if photon timing can be used to detect the presence of a population of as-of-yet unidentified MSPs. 

{\it The GCE in the non-Poisson limit.}  We focus on the scenario in which the GCE is dominated by 
a relatively small number of bright pulsars which lie just below the {\textit Fermi}-LAT threshold for identification 
as a point source.  This scenario may be thought of as the ``non-Poisson limit," 
because the resulting photon counts-in-pixels distribution will be non-Poisson: fluctuations in the photon counts will be dominated by the number of bright pulsars which happen to 
be in any pixel, rather than by Poisson fluctuations around the mean flux.  In particular, we will take as a benchmark the scenario originally discussed in~Ref.~\cite{Lee:2015fea} 
as an explanation for the observed preference of the GCE for a non-Poisson 
photon count distribution.\footnote{There 
have been many subsequent refinements in the application of non-Poisson template fitting to the GCE (see, 
for example,~\cite{Leane:2019xiy,Leane:2020pfc,Buschmann:2020adf,Mishra-Sharma:2021oxe}), 
but for the purposes of our study, this original scenario is a reasonable benchmark.}  In this scenario, the GCE is produced by $\sim 300 - 400$ unresolved MSPs.  The luminosity function for these pulsars 
(in the $1.9 - 12~\gev$ range) is strongly peaked at $\sim 1-2 \times 10^{-10} \cm^{-2} \s^{-1}$, with $\sim 100 - 150$ MSPs lying just below the point source  detection threshold.  Given a {\textit Fermi}-LAT exposure of $\sim 7 \times 10^{10} \cm^2 \s$, one would expect  {\textit Fermi}-LAT to have observed ${\cal O}(10)$ photons from each of these unresolved pulsars.  

One can estimate the number of photons per pixel arising from other sources from a photon count map of the GC (see, for example, Ref.~\cite{Lee:2015fea}).  One finds that, after the galactic plane 
is excluded, typical counts in a pixel  (we assume pixels corresponding to \texttt{healpix} $N_{\rm side} = 128$) in the $1.9 - 12~\gev$ energy range are $10 - 30$ photons, with $\gtrsim 50$ photons in only a few pixels.  
As benchmarks we will thus consider a 
pixel in which the expected number of photons over the {\textit Fermi}-LAT exposure time due to time-invariant 
sources is either 30 or  50, while the expected number of photons due to a single bright pulsar is 10 to 20.\footnote{It would 
be rare for two or more bright pulsars to lie in the same pixel, so we will ignore that possibility, as it 
will not affect our results significantly.} 
We will also consider scenarios in which the expected number of photons due to both a bright pulsar and time-invariant sources 
are considerably larger, as would be the case if one increased the energy range under consideration.

{\it Photon Emission Model.}  We will consider a simple pulsar emission model, in which the flux from the MSP is described by a periodic train of Gaussians, with the period $\tau$  and Gaussian pulse width, $w$.  As pulsar emission in gamma-rays is expected to be sharply peaked in time~\cite{Harding:1998wu}, we
will assume that $w \ll \tau$.  The total flux (photons per time) at time $t$ contributing to the pixel is then 
\begin{eqnarray}
\label{eq:rt}
    r(t; \theta)= \sum_{z = -\infty}^{z = \infty} r_s \exp\left[-\frac{(t  -  z \tau - t_0)^2}{2w^2}\right] \Theta(t, z, \tau) + r_b,
\end{eqnarray}
where  $\theta$ represents the model parameters, and
\begin{eqnarray}
\Theta(t,z, \tau) = \begin{cases}
  1  &  \text{if $z\tau + t_0  - \tau/2  < t < z\tau + t_0  + \tau/2$}  \\
  0 & \text{ otherwise}
\end{cases}
\end{eqnarray}
is a tophat function that cuts off the pulse so that it has no flux outside of a time period of length $\tau$.  In general, since $w$ is significantly smaller than $\tau$, truncating the tails of the pulse has negligible impact.  The sum over $z$ in Eq.~\ref{eq:rt} runs over all integers, but in practice we truncate the sum to the regime relevant to the finite time of the observations.  
The factors $r_s$ and $r_b$ control the amplitude of the MSP signal and the (time-independent) flux from background sources, respectively.  For convenience, we will treat the total expected number of 
photons observed over time $T$, from the MSP, $N_s$, and from backgrounds, $N_b$, as the free parameters rather than $r_s$ and $r_b$.  In other words, $N_b = r_bT$.  This simple model has five parameters: the expected number of photons over the {\textit Fermi} exposure 
from the pulsar ($N_s$) and from time-invariant sources ($N_b$), the period $\tau$ of the pulsar, 
the width of the pulse ($w$) and the phase ($t_0$).  We will refer to this set of parameters as $\theta_{\rm simple}$.

{\it Likelihood.  }  Assuming the distance between the detector and the emitters remains fixed (we will return to this assumption below), the above model enables us to calculate a likelihood for the observed photon times given $\theta_{\rm simple}$.  Consider a time interval $T$ divided into small subintervals $\Delta$ such that the number of photons received in each subinterval is either zero or one.  The  expectation value of the number of photons during $(t, t+\Delta)$ is then $\mu(t) = r(t) \Delta$.  The Poisson probability of detecting zero photons in the interval is
\begin{equation}
    P(0) = \exp(-r(t) \Delta)
\end{equation}
and the Poisson probability of detecting one photon is
\begin{equation}
    P(1) = (r\Delta) \exp\left(-r(t) \Delta \right).
\end{equation}
The total likelihood for the data vector, $\vec{d}$, consisting of ones and zeros representing photon detection/nondetection is then
\begin{eqnarray}
\mathcal{L}_{\rm simple}(\theta_{\rm simple} | \vec{d}) &=& \prod_i \left[ \delta_{0 d_i} P(0) + \delta_{1 d_i}P(1) \right] = \prod_i^{N_T - N} \exp(-r(t_i) \Delta) \prod_i^N  (r(t_i) \Delta) \exp\left(-r(t_i) \Delta \right) \nonumber \\
&=&  e^{-\int_T r(t_i) dt }\prod_i ^N  r(t_i) \Delta,
\label{eq:likelihood}
\end{eqnarray}
where $N$ is the number of intervals that have a detected photon, $N_T$ is the total number of intervals, $\delta$ is a Kronecker $\delta$, and the final product runs over all detected photons.  This likelihood was first derived in \citet{Gregory:1992}.

Of course, there are several effects that are not encapsulated by the simple model we have introduced.  For example, the 
motion of the Earth around the Sun will induce a perturbation to the photon arrival times as the distance between the detector and the emission source is modulated. 

The effect of some perturbations 
is to shift the arrival of the photons times in a completely deterministic 
and invertible way.  By this we mean that given some set of parameters, $\theta_{\rm pert}$ describing the perturbation, the perturbed detection time, $t'$ is related to the unperturbed time, $t$, by
\begin{eqnarray}
\label{eq:perturb_form}
    t' = f(t;\theta_{\rm pert}),
\end{eqnarray}
where $f$ is some function which can depend on $\theta_{\rm pert}$, and which is invertible.   
We restrict our consideration here to perturbations for which $dt'/dt = \dot f \sim 1$.

Perturbations of this type would include, for instance, perturbation to photon arrival times caused by Earth's orbit and by the orbit of {\textit Fermi} around the Earth.  
The effect of such perturbations
is to add or subtract some additional time delay to the photons; this perturbation could be inverted by simply subtracting off this time delay.  Moreover, for such perturbations, $\dot f -1 \sim {\cal O}(v/c) \ll 1$, where $v$ is a velocity scale characteristic of {\textit Fermi}, the Earth, the Sun, or the MSP.  As another example, note that MSPs can have a gamma ray pulse profile with two repeating pulses which are not equally spaced in time; such a perturbation is also of this type, with a constant shift in time for every other pulse. 
But not all perturbations to the photon arrival times can be written in this way, as we will discuss below. 

The key point is that if the perturbation can be written in this way, then it is in principle possible to solve for $t$ given the observed time $t'$ and the parameters $\theta_{\rm pert}$: $t = f^{-1}(t'; \theta_{\rm pert})$.

A key simplification now arises in the calculation of the likelihood.  In particular, since the perturbation can be exactly undone, we find 
\bea
{\cal L}(\theta_{\rm pert}, \theta_{\rm simple} | \vec{d}) &=& {\cal L}_{\rm simple}(\theta_{\rm simple} | f^{-1}(\vec{d};\theta_{\rm pert}) ). 
\label{eq:likelihood_relation}
\eea
In other words, the likelihood of obtaining data $\vec{d}$, given a model parameterized by $\theta_{\rm simple}$ 
and $\theta_{\rm pert}$, is the same as the likelihood of obtaining data $\vec{d}$ but with the effect of the 
perturbations on the data removed, given the simple pulsar model with parameters $\theta_{\rm simple}$.
Note that we have implicitly used the fact that $dt'/dt \sim  1$.  Generally, perturbations of this form will not 
only shift the photon arrival time, but will also rescale the probability density by a Jacobian factor; we consider 
perturbations for which this factor is close to 1.\footnote{More generally, perturbations of this form will rescale the 
width of pulse by the inverse of the Jacobian factor.  Assuming the pulse width is small compared to the time binning, 
the Jacobian factors will in any case cancel.}

In Eq.~\ref{eq:likelihood_relation}, every photon is assumed to have originated from 
the pulsar, and their arrival times are modified by removing the effect that the perturbation (parameterized 
by $\theta_{\rm pert}$) had.  Of course, many, if not most photons may have originated with the 
time-invariant background.  However, Eq.~\ref{eq:likelihood_relation} is nevertheless true:
since the background is time-invariant, the probability of a background photon arriving in one time 
bin is the same as in any other time bin. Thus, as long as perturbations to the MSP photons can be written as in 
Eq.~\ref{eq:perturb_form}, we can relate the likelihood of the data given these perturbations to the likelihood in their absence. 

{\it Measuring feasibility of detection.}
Our goal is now to determine if there is enough information in the {\textit Fermi}-LAT photon timing data 
to distinguish a preference for the model with a pulsar in the pixel over a model with no pulsar, 
and only a time-invariant background.  For this purpose, we will generate mock datasets with the expected statistical power of the {\textit Fermi} data.  To assess a preference for the model with a MSP over the model without a MSP we will utilize the Bayesian Information Criterion (BIC), which is given by
\bea
BIC &=& k \ln n - 2{\cal L}^{\rm max},
\eea
where $k$ is the number of parameters in the model, $n$ is the number of 
data points, and ${\cal L}^{\rm max}$ is the maximum likelihood of the data under the corresponding model.

We would like to compute the BIC for the model with a pulsar in the pixel 
($BIC^{\rm signal}$) and the BIC for the model with no pulsar in the pixel 
($BIC^{\rm bgd}$).  A smaller BIC implies a 
greater preference for the model, so we consider the quantity  
\bea
\Delta BIC &\equiv& BIC^{\rm signal} - BIC^{\rm bgd} ,
\nonumber\\
&=& \Delta k \ln n - 2\Delta {\cal L}^{\rm max} .
\eea
If $\Delta BIC \lesssim -10$, when summed over all pixels with anomalously high photon count, 
then {\textit Fermi}-LAT data has, at least in principle, enough information to distinguish 
the presence of a periodic source in those pixels.  Note, however, that since we expect 
about a  hundred such pixels hosting bright MSPs, $\Delta BIC$ averaged over all such pixels 
need only be slightly negative.

Unfortunately, we do not have a precise model for the perturbations to the periodic pulsar 
timing signal.  However, even without such a model, we can set an upper limit on $\Delta BIC$ with the following argument.  The maximum likelihood of the data over the full parameter space $(\theta_{\rm pert}, 
\theta_{\rm simple})$ is necessarily greater than or equal to the maximum likelihood of the same data 
over the restricted space $\theta_{\rm simple}$, with $\theta_{\rm pert}$ fixed to the true values.  But we argued above that when the perturbation can be exactly undone, the likelihood for the perturbed data can be related to the likelihood of unperturbed data under the simple model.  We therefore find
\bea
\mathcal{L}^{\rm signal}_{\rm max} = \max_{(\theta_{\rm simple}, \theta_{\rm perturb})} \mathcal{L}( \theta_{\rm simple}, \theta_{\rm perturb}|\vec{d}) > \max_{\theta_{\rm simple}}\mathcal{L}(\theta_{\rm simple} | f^{-1} (\vec{d} ; \theta_{\rm perturb}^{\rm true})).
\eea
For the time-invariant background model, perturbations to the photon times do not change the likelihood.  The maximum likelihood under the background-only model, $\mathcal{L}_{\rm bgd}^{\rm max}$,  can thus be computed by simply maximizing Eq.~\ref{eq:likelihood} over $r_b$ with $r_s$ fixed to zero:
\begin{eqnarray}
    \mathcal{L}^{\rm bgd}_{\rm max} = \max_{r_b}\mathcal{L}_{\rm simple}( r_b, r_s = 0 | \vec{d}) 
     =\max_{r_b}\mathcal{L}_{\rm simple}(r_b, r_s = 0 | f^{-1} (\vec{d} ; \theta_{\rm perturb}^{\rm true}) ).
\end{eqnarray}

We then have
\begin{eqnarray}
    \Delta \mathcal{L}^{\rm max} &=& \mathcal{L}^{\rm signal}_{\rm max} - \mathcal{L}^{\rm bgd}_{\rm max} \\
    &>& \max_{\theta_{\rm simple}}\mathcal{L}_{\rm simple}(\theta_{\rm simple}| f^{-1} (\vec{d} ; \theta_{\rm perturb}^{\rm true})) - \max_{r_b}\mathcal{L}_{\rm simple}(  r_b, r_s = 0 | f^{-1} (\vec{d} ; \theta_{\rm perturb}^{\rm true})) \\
    &\equiv& \Delta \mathcal{L}_{\rm simple}^{\rm max}.
\end{eqnarray}
This last term is just the difference in maximum likelihoods between the simple model and background-only model, evaluated on data with no perturbations.  The arguments above demonstrate that this difference sets a lower limit on the true $\Delta \mathcal{L}^{\rm max}$.  

We constrain $\Delta \mathcal{L}^{\rm max}_{\rm simple}$ by simulating unperturbed data, and finding the maximum likelihood over the $\theta_{\rm simple}$ parameter space and over the background-only parameter space.  Finally, our constraint on $\Delta \mathcal{L}_{\rm simple}^{\rm max}$ translates into a limit on $\Delta {\rm BIC}$:
\bea
\Delta BIC &<& \Delta k \ln n - 2\Delta {\cal L}_{\rm simple}^{\rm max} .
\label{eq:DeltaBIC_min}
\eea

{\it Results.}  We generate mock data sets from the simple  model, fixing the true parameters to $(\tau, w, t_0) = (5\,{\rm ms}, 0.5\,{\rm ms}, 0)$.  
A ratio $w/\tau \sim 0.1$ is typical of known MSP gamma ray pulse profiles~\cite{Benli:2020qql}.
We set the time resolution  to $5\times 10^{-4}\,{\rm s}$, the same as the pulse width, and easily achievable given the performance of {\textit Fermi}.  
{\textit Fermi}'s time resolution is significantly better than $5\times 10^{-4}$s, and in principle finer binning could be used. Our choice of coarse binning means that the pulses are highly under-sampled.  
The motivation for this choice is to
reduce our sensitivity to perturbations to pulse shape, etc.  We explore the sensitivity of our results to these choices in the Appendix.  We set the total observation time to $10^{8}\,{\rm s}$.  We will consider several choices of $N_s$ and $N_b$.

\begin{figure}
    \includegraphics[scale = 0.5]
    {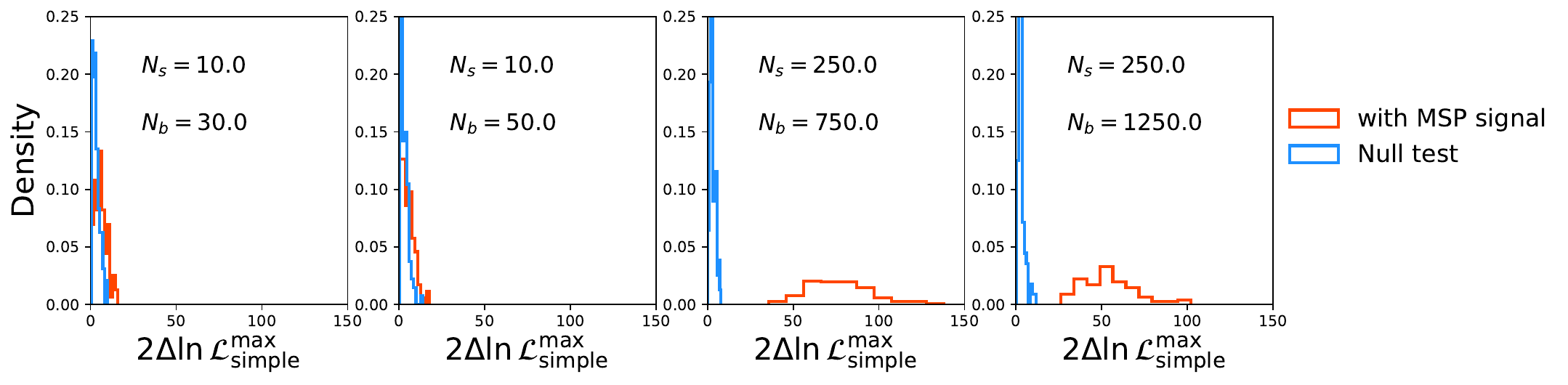}
    \caption{Histograms of $2\Delta \ln \mathcal{L}^{\rm max}_{\rm simple}$ from the mock data analysis.   In each case, we have generated 100 mock data sets at the true parameter values indicated in the text, and for the values of $N_b, N_s$ indicated at the top of each panel (red).  The blue histograms are similar, but describe 
    the case in which the parameter values used to generate the mock data are for the background-only model, with $N_s +N_b$ held fixed.}   
    \label{fig:DeltaLnL}
\end{figure}

For a given mock data set, we maximize the likelihood over $\theta_{\rm simple}$ using the basin hopping algorithm \citep{BasinHopping}.  We also compute the maximum likelihood for the background-only model, varying only $r_b$.  The difference between these maximum likelihood values is then $\Delta \mathcal{L}^{\rm max}_{\rm simple}$.  Finally, we repeat this process many times to build a distribution for $2\Delta {\cal L}_{\rm simple}^{\rm max}$.  The condition 
$\Delta k \ln n < 2\Delta {\cal L}_{\rm simple}^{\rm max}$ will then allow us to estimate 
the maximum number of parameters $\theta_{\rm pert}$ which can perturb our simple periodic model, 
while still guaranteeing that there is sufficient information in the data to determine the presence of periodic sources.  Note that since we are only setting a lower limit on $\Delta \ln \mathcal{L}^{\rm max}$, it is possible that a detection could be achieved with more than this number of parameters.  Our results can therefore be viewed as conservative.
We plot the distributions of $2\Delta {\cal L}_{\rm simple}^{\rm max}$ in Figure~\ref{fig:DeltaLnL} (red histograms) for 
data generated assuming $(N_s, N_b) = (10,30)$ (far left), $(10,50)$ (middle left), 
$(250,750)$ (middle right) and $(250,1250)$ (right).  
In each panel, we also plot the distribution of  
$2\Delta {\cal L}_{\rm simple}^{\rm max}$ when the true data includes no signal contribution, but is background-only  with 
an expected number of photons given by the value of $N_s + N_b$ used in each panel (blue histograms, labelled `Null test').
For $N_s = 10$ (and $N_b =30$ or $50$) we find that the distribution of 
$2\Delta {\cal L}_{\rm simple}^{\rm max}$ for a pixel containing a bright MSP is essentially indistinguishable 
from one with only time-invariant backgrounds, indicating that there is clearly not enough information 
in the timing data to detect evidence of the periodic MSP (left two panels of Fig.~\ref{fig:DeltaLnL}).    However, we may consider the possibility of 
increasing the energy range of the photons included in our analysis in order to increase the number of photons arising 
from a periodic source.  It was estimated above that an MSP lying just below the point source detection threshold 
would produce $\sim 10$ photons in the energy range $2-12$~\gev.  However, {\textit Fermi} can detect photons up to 
an order of magnitude lower in energy.  Since an MSP gamma-ray energy spectrum scales roughly as $E^{-1.5}$~\cite{Cholis:2014noa}, 
one might expect that up to $N_s \sim 250$ could be reasonably obtained, for an MSP just below the 
point source detection threshold, with current {\textit Fermi} exposure, given an energy window above the {\textit Fermi} energy threshold. If we assume that this increase in the energy range increases $N_b$ proportionally, then our most optimistic benchmarks would be $(N_s,N_b$) = $(250,750)$ or $(250,1250)$.

If $N_s =250$ (for either $N_b =750$ or $1250$) the distribution is 
peaked at $2\Delta \ln {\cal L} \sim 70$, implying that if the number of parameters needed to describe the pulsar model is
$\Delta k \lesssim 70/\ln(1500) \approx 10$, then there will be sufficient information in the data to detect the periodic MSP signal (right two panels of Fig.~\ref{fig:DeltaLnL}).  Beyond $N_b$ 
and the Gaussian width (which does not affect the likelihood when the time binning is sufficiently 
coarse), the simple pulsar model has 3 parameters, implying that perturbations to the arrival 
times described by less than roughly $7$ parameters are still guaranteed adequate detection prospects.   

In Figure~\ref{fig:variations} in the Appendix, we repeat this analysis for a few other choices of $N_s$, $N_b$, 
$\Delta T$ and $w$, in order 
to determine how the distribution of $2\Delta {\cal L}_{\rm simple}^{\rm max}$ changes with these parameters.  We 
see that increasing $N_b$ has a relatively mild effect on the likelihood, compared to increasing 
$N_s$.  Evidently, our ability to detect the presence of a periodic source is limited less by the size of the 
time-invariant background than by the necessity of having enough periodic photons to constrain the model parameters.
We also see that reducing the size of the pulse width, in comparison to the period, tends to lead to greater 
discriminating power.  Thus, the presence of pulsars with smaller gamma-ray pulse widths than our benchmark 
assumption can lead to improved sensitivity.

\textit{Other perturbations.}
Some perturbations to photon arrival times may not be expressible in a form like Eq.~\ref{eq:perturb_form}.    One such perturbation, for example, is a possible spin down of the pulsar.  Spin down will change the times of the pulses, but it will also change their widths.  
More generally, beyond changes to the arrival time of a pulse, perturbations can change the pulse width, or the expected number of photons per pulse.  Since we use time bins that are comparable in width to the pulses, changes to the pulse widths over the observation time would have to be ${\cal O}(1)$ in order to significantly impact our results.  But we are justified in 
assuming that the effects of spin down cannot  be large on the timescale of {\textit Fermi} observation.  Given that the age of a typical MSP is ${\cal O}(10^9)\yr$, it is unlikely 
that a pulsar from which gamma rays can be detected with a ${\cal O}(10^{-3})\s$ period today, could have either a luminosity in gamma-rays which is decreasing rapidly, or 
a width which is increasing rapidly with respect to the bin size, over the period of observation.  

For radio pulsars, an additional source of perturbations to pulse timing is propagation of the pulses through plasma, leading to a pulse delay that is related to the dispersion measure.  This effect would not be of the form Eq.~\ref{eq:perturb_form} because of its stochastic nature: the dispersion measure can vary with time as the relative positions of the Earth, plasma, and MSP change.  For gamma-rays, however, such propagation effects are expected to be negligible \citep{Fermi-LAT:2022wah}.

{\it Discussion.}
We have considered a strategy for testing if a sizeable contribution to the Galactic Center GeV excess arises from a population of 
bright millisecond pulsars, which each produce several photons seen by {\textit Fermi}-LAT.  Such a population of MSPs may be responsible for 
non-Poisson fluctuations in the photon counts in pixels found in some recent analyses of the GC excess.  Our strategy is to use the fact 
that the timing resolution of {\textit Fermi}-LAT (${\cal O}(10^{-6})\s$) is much smaller than period of these sources, suggesting that {\textit Fermi}-LAT 
data may contain enough information to detect the presence of periodic sources with a high degree of confidence.

We have found that if a typical bright pulsar produces $\sim 250$ photons seen by {\textit Fermi}-LAT within the energy range of interest, then 
there is, in principle, enough information in the photon timing data to identify the presence of $\sim 1\,{\rm ms}$ periodicity in the photon arrival times, even if these times are 
perturbed via a perturbation model with $\sim 7$ free parameters.  

Our analysis has essentially been a proof of principle.  It would be interesting to apply this type of analysis to current {\textit Fermi}-LAT data.  
A first step would be the development of a robust pulsar gamma-ray emission model, including a detailed parameterization of the effects 
which could perturb the timing of gamma ray emission.

In this analysis, we have only considered the timing of photons in a pixel, not the detailed energy spectrum.  Since the energy spectrum of 
pulsar gamma-ray emission will differ from that of other astrophysical backgrounds, additional information may be contained in correlations 
between the photon energy and photon arrival times.  A more detailed study of this possibility would be an interesting topic for future work.

{\bf Acknowledgements.}  We are grateful to Kevork Abazajian, Katharena Christy, Addy Evans, and Dan Hooper for useful discussions.  
For facilitating portions of this research, the JK wishes to acknowledge the Center for Theoretical Underground Physics and Related Areas (CETUP*), The Institute for Underground Science at Sanford Underground Research Facility (SURF), and the South Dakota Science and Technology Authority for hospitality and financial support, as well as for providing a stimulating environment.
JK is supported in part by DOE grant DE-SC0010504.  

\bibliography{thebib}

\section*{Appendix}

\begin{figure}[!h]
    \includegraphics[scale = 0.5]{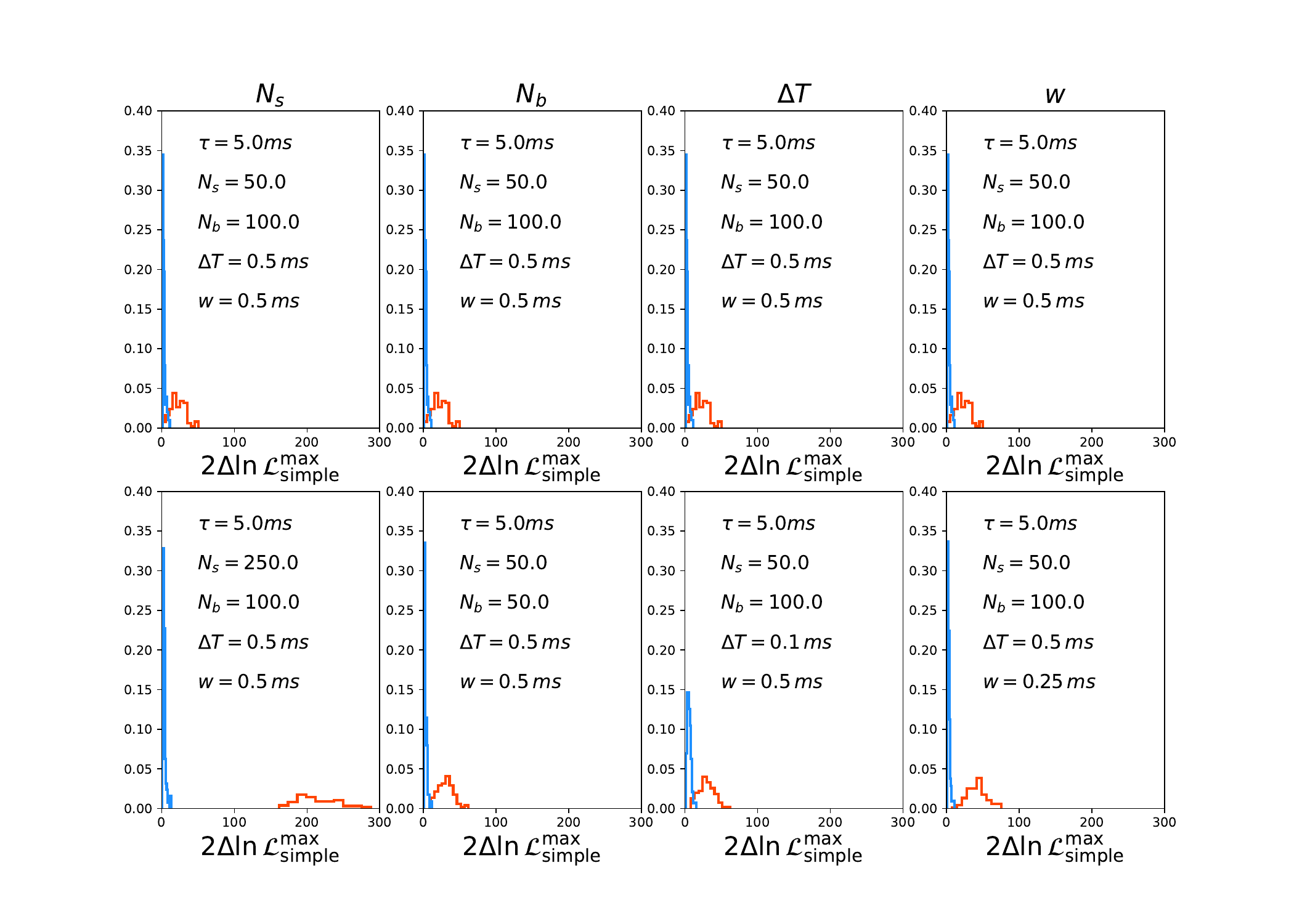}
    \caption{Same as Fig.~\ref{fig:DeltaLnL}, but showing more variations in parameter choices.  The top panels are all the same, with parameter choices indicated.  The bottom panels vary a single parameter relative to the top panels, with the choice of parameter indicated above each column.}
    \label{fig:variations}
\end{figure}

\end{document}